\documentclass[UTF8]{article}
\date{} 
\usepackage{ctex}
\usepackage{amsmath, amssymb, amsthm}
\usepackage{booktabs}
\usepackage{caption}
\usepackage{multirow}

\usepackage{graphicx}
\usepackage{listings}
\usepackage{xcolor}
\usepackage{natbib}
\usepackage{enumitem}
\usepackage{titlesec} 
\usepackage{array}
\newcolumntype{L}{>{\ttfamily}l} 
\usepackage{natbib}      
\usepackage{tikz}
\usepackage{titlesec}
\usepackage{fancyvrb} 
\usepackage{diagbox}

\usepackage{times}  

\usepackage{listings}  
\usepackage{fontspec}
\usepackage{hyperref}    
\hypersetup{
	colorlinks=true,     
	linkcolor=black,      
	citecolor=black,       
	urlcolor=cyan        
}

\usepackage{xeCJK}
\usepackage[top=2.5cm, bottom=2.5cm, left=2.5cm, right=2.5cm]{geometry}
\usepackage{fancyhdr}
\fancypagestyle{FirstPage}{
	\fancyhf{}
	\fancyhead[L]{\normalsize\textit{doi \quad 编号}} 
}

\fancypagestyle{OtherPages}{
	\fancyhf{}
	
}

\makeatletter

\renewcommand{\footnote}{}    

\def\@makefnmark{}              

\titleformat{\section}[block] 
{\normalfont\Large\bfseries\songti}
{\thesection} 
{0em} 
{}
\titlespacing*{\section}{0pt}{1.5ex}{1ex} 

\usepackage{natbib}
\setcitestyle{super,square,sort&compress}

\newenvironment{enabstract}{%
	\textbf{Abstract:} 
	\noindent\ignorespaces 
}{%
	\par\vspace{0.5em} 
}

\title{A Quantization-Aware Training Based Lightweight Method for Neural Distinguishers
}
\author
	{
		\zihao{4}Xiong Guangwei\thanks{Xiong Guangwei(2001-), Master student, major research interests: intelligent information processing. Email: GuangweiX@hotmail.com; }，
		\zihao{4}Wang Linyuan\thanks{Wang Linyuan(1985-), Associate Professor, Ph.D., major research interests: mathematical foundation of artificial intelligence, neural network architecture design and optimization. Email: wanglinyuanwly@163.com；}，
        \zihao{4}Zheng Zhizhong\thanks{Zheng Zhizhong(1981-), Associate Professor, Ph.D., major research interests: optimal transport theory, intelligent cryptanalysis. Email: zhengzz81@163.com；}，\\
        \zihao{4}Hou Senbao\thanks{Hou Senbao(1995-), Ph.D. candidate, major research interests: large model compression, computer-aided diagnosis, pattern recognition. Email: hsb378093739@163.com；}，
        \zihao{4}Yan Bin\thanks{Yan Bin(1976-), Professor, Ph.D., major research interests: fundamental theory of artificial intelligence, intelligent information processing, hybrid intelligence for brain-computer interaction. Email: ybspace@hotmail.com}，\\
        \zihao{5}(Henan Key Laboratory of Imaging and Intelligent Processing, Information Engineering University, \\
        \zihao{5}Zhengzhou Henan 450001, China)} 

\makeatother
\makeatletter
\def\thanks#1{\protected@xdef\@thanks{\@thanks
        \protect\footnotetext{#1}}}
\makeatother

\begin{document}
\maketitle
\thispagestyle{FirstPage} 



\begin{enabstract}
	\noindent 
	In 2019, Gohr pioneered the application of deep neural networks to differential cryptanalysis, developing DNN-based neural distinguisher classifiers to analyze the SPECK lightweight block cipher. Unlike traditional differential analysis, which relies on Boolean operations on 0-1 sequences, neural distinguishers extract continuous features, introducing 32-bit multiplications operations that increase complexity and potential redundancy. This study proposes a lightweight neural distinguisher based on quantization-aware training. Leveraging learnable step-size quantization, the model's weights are quantized to 1.58 bits, enabling the replacement of all convolutional multiplication operations with Boolean logic. Additionally, the ReLU activation function is reimplemented as a comparison-based indicator function. This transforms the original 32-bit multiplication-dependent architecture into a lightweight structure composed solely of Boolean operations, additions, and indicator functions. Experimental results confirm significant computational complexity reduction. Owing to a high proportion of zero-valued weights, the total operations amount to just 13.9\% of Gohr's model. Critically, the most costly 32-bit multiplications are eliminated, with classification accuracy dropping by only 2.87\%. When applied exclusively to the initial convolutional layer, the 128 1-by-1 convolutions are replaced with 4 Boolean operations on 16-bit sequences, incurring a negligible 0.3\% accuracy loss..\\
	\hangindent=2em \noindent\hspace*{2em}\textbf{Keywords:} SPECK Lightweight Block Cipher; Neural Distinguisher; Quantization-aware Training; Lightweight Methods \\
\end{enabstract}

\section{\quad Introduction} 
Block ciphers have been widely adopted in domains such as finance, communications, and the Internet of Things due to their high encryption/decryption efficiency and relatively low hardware implementation cost \cite{1}. In 1990, Adi Shamir et al\cite{2}. first publicly proposed differential cryptanalysis. By exploiting the statistical propagation behavior of differences, this method constructs high-probability “differential trails” (i.e., paths along which a plaintext difference is transformed into a specific ciphertext difference after multiple rounds), and combines them with ciphertext-difference statistics obtained from chosen plaintext pairs to effectively reduce the key search space.

In 2019, Gohr \cite{3} applied neural distinguishers(ND) to differential cryptanalysis in order to capture distributional characteristics of encrypted ciphertext pairs and thereby determine whether a given input difference corresponds to a real differential or a random one, providing effective assistance for differential cryptanalysis. Gohr \cite{3} was also the first to incorporate residual blocks into the design of ND; by introducing skip connections, the vanishing-gradient issue in deep networks can be alleviated. On the SPECK32/64 cipher, this approach enabled efficient classification of ciphertext pairs generated by seven rounds of encryption, and further supported an 11-round SPECK32/64 last-round subkey recovery attack, laying the foundation for applying ND in differential cryptanalysis.

With respect to architectural design and optimization of ND, Bellini et al. \cite{4} focused on improving generality and proposed a generic ND based on dilated convolutions. By introducing dilation factors in convolutional layers and stacking parallel branches with different dilation rates, their model can be applied to ciphertext pairs generated by multiple ciphers without additional, cipher-specific redesign. Bao et al. \cite{5} incorporated dense connections and squeeze-and-excitation (SE) blocks into ND architectures, thereby broadening the space of candidate network designs and enabling a more systematic exploration of alternative structures.

With respect to input design and optimization of ND, Chen et al. \cite{6} proposed a ND that jointly classifies multiple ciphertext pairs. Their method extends the conventional single-pair input setting to a parallel multi-pair input, enabling the model to process k ciphertext pairs simultaneously and to extract derived features across pairs. As a result, the classification accuracy on the SPECK32/64 7-round dataset is further improved. Hou et al. \cite{7} designed a dedicated ND for the SIMON32/64 cipher and systematically investigated the relationship between input-difference patterns and classification performance; based on these findings, they achieved a SIMON32/64 key-recovery attack that is more efficient than traditional cryptanalytic approaches. From the perspective of input representation, Liu et al. \cite{8} integrated domain knowledge from differential cryptanalysis with deep-learning-based data representation and proposed a two-dimensional “non-real” input generation method. Building on this idea and incorporating multi-ciphertext-pair analysis, they further developed two concrete input construction schemes, namely multi-round multi-concatenated pairs and multi-round multi-concatenated differences.

The above studies \cite{4,5,6,7,8} have substantially improved the classification performance of ND. However, these performance gains are often accompanied by a notable increase in model complexity, which has motivated growing interest in lightweight ND. Zhang et al. \cite{9} constructed ND for SPECK32/64 and SIMON32/64 by incorporating an Inception module and knowledge distillation. The Inception module employs parallel branches with $1\times 1$ convolutions, $3\times 3$ convolutions, and pooling to capture multi-scale features and expand the receptive field, while knowledge distillation facilitates model compression and accelerates inference. Bacuiet et al. \cite{10} applied the lottery ticket hypothesis to ND for SPECK; using a pruning criterion based on zero mean activation, they obtained ND that are either smaller in size or superior in performance. Liu et al. \cite{11} addressed cost optimization in differential neural cryptanalysis by proposing a new ND model informed by traditional cryptanalytic techniques and theoretical analysis. By reducing the training sample size and introducing depthwise separable convolutions, their approach decreases both training computation and parameter count while maintaining high accuracy. Ebrahimi et al. \cite{12} proposed a partial-differential ND, which relaxes the requirement of training on the full block of bits. Their model is trained using only a subset of key bits and can still distinguish whether a ciphertext difference is produced by the cipher or arises randomly, thereby significantly reducing the amount of required input features.

We note that in the above lightweight-model studies \cite{9,10,11,12}, the feature extraction process in ND is predominantly continuous, and the resulting representations are also continuous-valued. In contrast, the block ciphers considered in this work are inherently based on discrete operations, and the inputs to the ciphertext classification task are likewise discrete. As pointed out in \cite{13}, such a mismatch in feature representation during training may hinder ND from accurately capturing key information in ciphertext data, thereby degrading classification accuracy. Moreover, ND based on continuous features typically require a large number of 32-bit multiplications, which are more computationally expensive and may be redundant when compared with Boolean logic operations on binary (0/1) sequences. To address these issues, we adopt quantization-aware training to quantize the ND to 1-bit precision, then replace multiplications with Boolean operations, and substitute the ReLU nonlinearity with an indicator function based on comparison operations. This yields a lightweight ND and significantly reduces the computational cost of the original model.

\section{\quad Method}
To address the high computational cost of ND due to the extensive use of 32-bit multiplications, this paper proposes a lightweighting method based on quantization-aware training. The proposed procedure is as follows: we apply quantization-aware training to quantize the weights of Gohr's ND to 1.58-bit precision with values in (${0,\pm1}$); we then replace multiplication operations with Boolean logic, and substitute the ReLU nonlinearity with a comparison-based indicator function. In this way, a lightweight ND is obtained. Section 1.1 reviews the architecture of Gohr ND and the functionality of its components, while Section 1.2 provides a detailed description of the proposed lightweighting method.

\subsection{Gohr's Neural Distinguisher}
Gohr proposed several differential-cryptanalysis approaches for SPECK, where the central task is to distinguish real ciphertext pairs from random ciphertext pairs. A real ciphertext pair is obtained by encrypting a plaintext pair with a fixed input difference, whereas a random ciphertext pair is generated by encrypting a randomly sampled plaintext pair whose difference is random. Gohr compared the performance of traditional differential distinguishers and ND on $5–8$ rounds of SPECK32/64 and observed that ND achieve superior performance.

A traditional differential distinguisher is typically constructed under the Markov assumption. Given a ciphertext pair \((C, C')\), it first computes the output difference \(\Delta_{\text{out}} = C \oplus C'\), and then evaluates the transition probability $p$ of \(\Delta_{\text{in}} \to \Delta_{\text{out}}\). If \(\text{DDT}(\Delta_{\text{in}} \to \Delta_{\text{out}}) > \frac{1}{2^{32} - 1}\), the pair is classified as a real ciphertext pair; otherwise, it is classified as a random ciphertext pair.

\begin{figure}
    \centering 
    \includegraphics[height=6cm,width=8cm]{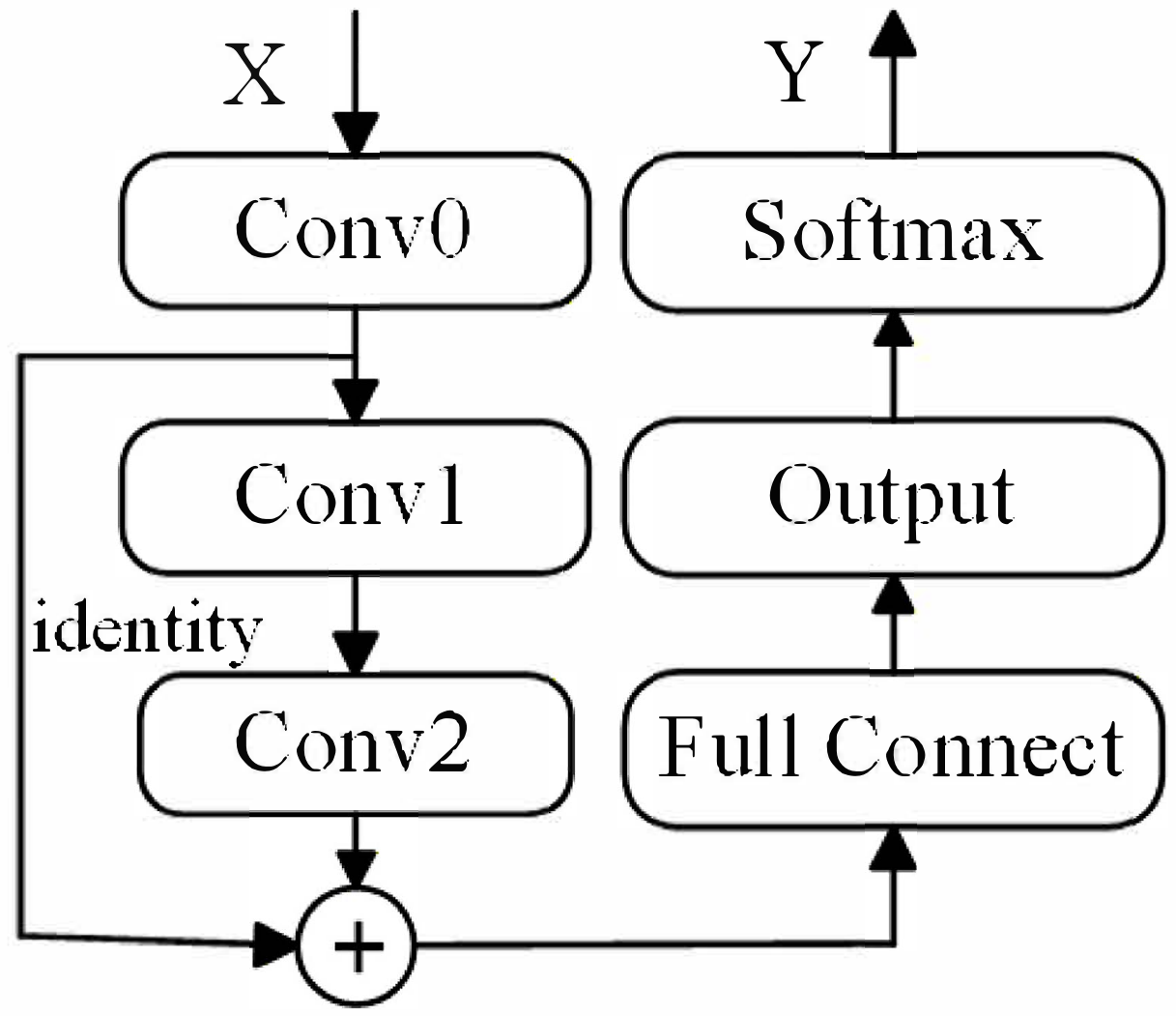}
    \caption{Residual-Block-Based ND} \label{figure1}
\end{figure}

Gohr's ND adopts a residual network as its core architecture and consists of multiple components, including an input layer, several residual blocks, and a prediction head. The input layer represents each SPECK32/64 ciphertext pair—composed of four 16-bit words—as a 4×16 matrix, and then uses an initial convolutional layer to extract features into a 32×16 feature map. The residual backbone stacks multiple convolutional blocks (each including 3×3 convolutions, batch normalization, and ReLU nonlinear activations) together with skip connections to facilitate gradient propagation; both the input and output of the residual backbone are 32×16 feature maps. The prediction head is responsible for the final classification of ciphertext pairs. The detailed architecture is illustrated in Fig.\ref{figure1}. For each ciphertext pair, the ND outputs a probability estimate in the range [0,1].

In terms of data generation, Gohr uses a random number generator to sample uniformly distributed keys and to generate plaintext pairs with a specified input difference. Real samples are obtained by encrypting such plaintext pairs under SPECK32/64 for a given number of rounds. Random samples are produced by replacing the second plaintext in each pair with a uniformly random value before encryption.

During the training phase, the ND first converts ciphertext pairs into a 4×16 feature matrix $(C_l, C_r, C_l', C_r')$ as input.The initial convolutional layer adopts 32 channels with 1×1 convolutional kernels, mapping the input features derived from $(C_l, C_r, C_l', C_r')$ into 32×16 features.The core feature extraction component is the residual block, which enhances gradient propagation via multiple convolutional blocks and skip connections.Both its input and output are 32×16 features, and all employed convolutional kernels are 3×3 in size with a padding size and stride of 1.The prediction head adopts a densely connected structure without pooling layers.It first flattens the output of the residual block into a 512×1 vector, gradually compresses the features through three fully connected layers, and finally outputs a probability estimate in the range of 0–1 using the Sigmoid function and Softmax.If the estimate is $\geq0.505$, the input is determined as a genuine ciphertext pair; otherwise, it is classified as a random ciphertext pair.

\subsection{A Lightweight Method for Neural Distinguishers
Based on Quantization-Aware Training}
From the perspective of the ciphertext classification task, the input consists of 0–1 binary sequences and the output is a 0–1 binary classification label. When deep neural networks are used as distinguishers, they mainly extract continuous transformation features and involve a large number of 32-bit multiplications. Compared with the Boolean operations adopted in traditional cryptanalysis, this significantly increases computational overhead and may introduce considerable redundancy.
Quantizing the weights of the ND—especially quantizing them to 1‑bit—can substantially reduce computational overhead. Therefore, the core research idea of this paper is to apply quantization techniques to quantize the weights of the ND to 1‑bit while preserving accuracy as much as possible.
The lightweight method proposed in this paper mainly includes two components: weight quantization of the ND and operation simplification. The specific implementation methods of these two parts are introduced in detail below.
\subsubsection{1-bit quantization}
The quantization-aware training method \cite{14} is constructed based on the Learned Step Size Quantization (LSQ) framework \cite{15}, aiming to solve the training stability problem of low-bit models through a staged training strategy. This method integrates the quantization process with the training process, dynamically optimizes low-precision mapping via learnable quantization step-size parameters, and designs a temporal control mechanism to mitigate the accumulation of gradient approximation errors.
The core of the LSQ method is to realize the mapping from full-precision weights to low-bit integers through layer-wise trainable step-size parameters $\Delta$: a separate step-size parameter $\Delta$ is assigned to each layer of the Gohr's ND, and $\Delta$ participates in training synchronously with the weights of the ND.$\Delta$ is dynamically adjusted via gradient descent, so that the quantization step size of each layer adapts to the weight distribution characteristics of that layer.During forward propagation, the weight $w$ is quantized according to the function shown in Equation (\ref {eq:1}):
\begin{equation}\label{eq:1}
    \hat{w}_k = \operatorname{clip}\left( \operatorname{round}\left( \frac{w}{\Delta} \right) \times \Delta,\ -\Delta \times 2^{k-1},\ \Delta \times \left( 2^{k-1} - 1 \right) \right)
\end{equation}

Here, $\Delta$ acts as the learnable step‑size parameter, and $k$ denotes the preset quantization bit‑width.They are dynamically adjusted via gradient descent during training to minimize the accuracy loss caused by the quantization operation.
In this paper, we set $k=1$, so the quantization formula for weight w is given in Equation (\ref {eq:2}).
\begin{equation}\label{eq:2}
    \hat{w}_1 = \operatorname{clip}\left( \operatorname{round}\left( \frac{w}{\Delta} \right) \times \Delta,\ -\Delta,\ \Delta \right)
\end{equation}

The round function in the quantization process is a nonlinear step function whose derivative is identically zero at all non-integer input points. This implies that gradient information vanishes completely in such regions, making it impossible to transmit the gradient signals required for parameter update.
LSQ introduces the gradient approximation strategy known as the Straight-Through Estimator (STE). By decoupling the forward and backward propagation processes of the quantization operation, it enables effective gradient propagation.
In the forward propagation phase, LSQ preserves the nonlinear characteristics of the quantization operation and maps values to discrete quantization levels via the round function.In the backward propagation phase, however, LSQ adopts a gradient approximation strategy: it actively ignores both the step nonlinearity of the round function and the clipping nonlinearity of the clip function, and approximates the gradient function of the entire quantization operation as an identity mapping.
When computing the gradient of the loss function with respect to the full-precision weights, the gradient of the quantized weight $\hat{w}_1$ is directly propagated to the corresponding full-precision weight $w$, as shown in Equation (\ref {eq:3}).

\begin{equation}\label{eq:3}
    \frac{\partial L}{\partial w} \approx \frac{\partial L}{\partial \hat{w}_k}
\end{equation}

Through this design, gradients can successfully propagate through the quantization operation layer, breaking through the gradient interruption bottleneck caused by the non-differentiability of the round function. It provides continuous and effective gradient information for the iterative update of full-precision weights, enabling stable parameter optimization of the quantized model during training.

During quantization, to avoid significant accuracy loss and drawing on successful experiences from multiple studies, the method proposed in this paper does not implement strict 1‑bit quantization in the absolute sense.Instead, the weight values of the Gohr's ND are set to three cases: positive values (+1), negative values (−1), and zero (0).Accordingly, all mentions of “1‑bit quantization” in this paper specifically refer to this ternary quantization scenario. Calculated based on information entropy, its equivalent quantization precision is 1.58 bits, which is also directly referred to as 1.58‑bit quantization in many literatures.

\subsubsection{Operation Simplification of the Quantized Neural Distinguisher}

After quantization, the weights of the ND take only three discrete values, namely $+1$、$-1$ and $0$. From an operational perspective, the core procedure of a standard convolution can be described as follows. Let the input feature map be $X\in\mathbb{R}^{C\times H\times W}$, where $C$ denotes the number of channels and $H$ and $W$ are the height and width of the input, respectively. Let the convolution kernel be $W\in\mathbb{R}^{C\times K\times K}$, where $K$ is the kernel size. The kernel slides over the spatial dimensions of the input feature map with stride $s$. For each spatial location within a sliding window $X_{c,i:i+K-1,j:j+K-1}$, where $c$ is the channel index and $i,j$ denote the starting spatial coordinates, the kernel weights in each channel are multiplied element-wise with the corresponding local patch of the input feature map. The resulting products are then summed across all channels, yielding the value of the output feature map $Y\in\mathbb{R}^{H'\times W'}$ at position $(i',j')$. The formal definition is given in (\ref{eq:4}), where $H',W'$ denote the height and width of the output feature map, respectively.

\begin{equation}\label{eq:4}
    Y(i',j') = \sum_{c=1}^C \sum_{k_1=1}^K \sum_{k_2=1}^K W(c,k_1,k_2) \cdot X(c,i+k_1-1,j+k_2-1)
\end{equation}

In the simplified ND, the sign of the output is determined solely by the accumulated sum of the inputs associated with the two effective (non-zero) weight types. Since the product of a zero-valued weight and any input element is always zero, zero-valued weights make no contribution to the accumulation and can therefore be removed directly. The sets P and N are defined as in (\ref{eq:5}):

\begin{equation}\label{eq:5}
\begin{aligned}
    P &= \{(c, k_1, k_2) \mid W(c, k_1, k_2) = +1\} \\
    N &= \{(c, k_1, k_2) \mid W(c, k_1, k_2) = -1\}
\end{aligned}
\end{equation}

Here, the set P denotes all input-feature positions whose corresponding weights are $+1$, while the set N denotes all input-feature positions whose corresponding weights are -1. At each position associated with a non-zero weight, we perform a single Boolean “AND” operation between the (non-zero) weight indicator and the input bit, and then sum the resulting AND outputs. The accumulated input sum is computed as in (\ref{eq:6}):

\begin{equation}\label{eq:6}
\begin{aligned}
    S_P &= \sum_{(c, k_1, k_2) \in P} X(c, i+k_1-1, j+k_2-1) \cdot W(c, k_1, k_2)\\
    S_N &= \sum_{(c, k_1, k_2) \in N} X(c, i+k_1-1, j+k_2-1) \cdot W(c, k_1, k_2)
\end{aligned}
\end{equation}

The output of the original ND's convolutional layers is typically obtained after applying an activation function. In the simplified method proposed in this paper, however, the output $Y$ is produced by an indicator function $I(\cdot)$, defined as follows: if the value of the expression inside the function is greater than 0, then $Y=1$; otherwise (i.e., if it is less than or equal to 0), $Y=0$. Accordingly, when the accumulated sum of the inputs associated with positive weights +1 is greater than that associated with negative weights -1, the output Y equals 1; otherwise, Y equals 0, as shown in (\ref{eq:7}).

\begin{equation}\label{eq:7}
    Y = I(S_P - S_N > 0)
\end{equation}

In the proposed simplification, both the inputs and the quantized weights of the ND take binary values in ${0,1}$, so the original convolution and nonlinear functions can be replaced by simple operations. Specifically, we first apply a quantization preprocessing step to the network weights, quantizing them to 1.58-bit precision. We then replace multiplication operations in the ND with Boolean operations, and substitute the ReLU nonlinearity in the network with a comparison-based indicator function. These modifications ultimately yield a lightweight ND.

\section{\quad Experiment}
This section simplifies Gohr's ND and evaluates the performance of the simplified model. All experiments were conducted on a workstation equipped with an NVIDIA GeForce RTX 3070 GPU, and the implementation was based on Python 3.9 and PyTorch 1.13.0.

This section evaluates the classification accuracy of the ND. Following the procedure in \cite{3}, we generate training and validation data using a random number generator to obtain uniformly distributed keys $K$, plaintext pairs $P$ with a fixed input difference $\Delta = 0\text{x}0040/0000$, and binarized real/random labels $Y$. For the generation of training or validation data for 6-round $\text{SPECK32/64}$, if $Y$ indicates a real sample, the plaintext pair $P_i$ is encrypted for $k$ rounds; if $Y$ indicates a random sample, the second plaintext in the pair is replaced with a newly generated random plaintext before encryption. Each class contains 5 million samples, resulting in 10 million samples in total. To mitigate randomness in a single run, we independently repeat the experiment 10 times and report the average classification accuracy over these runs.

To improve experimental accuracy, we follow the method in \cite{6} and extend the one-dimensional convolution to a two-dimensional convolution. Specifically, samples in the above dataset with the same label are grouped into sets of eight, thereby expanding the data from a one-dimensional format to a two-dimensional format. As a result, the input is reshaped from a 4×16 feature map to a 4×16×8 feature map.

Gohr ND consists of an initial convolutional layer, a feature-extraction module, a prediction head, and an output layer. In the following, we simplify the architecture of Gohr ND layer by layer. The numbers of multiplications and additions reported in this paper are obtained by counting the total number of multiplication and addition operations performed during inference. Multiplications mainly arise from element-wise products between convolutional kernels and their corresponding convolution inputs, while additions mainly refer to the summation of these products after the element-wise multiplications.

In the simplified implementation, the number of Boolean operations is given by the product of the number of non-zero weights and the output feature dimension. The number of additions is given by the product of ((\text{number of non-zero weights} - \text{number of output channels})) and the output feature dimension. The number of indicator-function evaluations is given by the product of the number of output channels and the output feature dimension.

In the quantized Gohr's ND, the initial convolutional layer has 4 input channels, and among the output channels, 4 channels contain non-zero weights. The distribution of these non-zero weights is reported in Table \ref{tab:1}. In the original Gohr ND, the initial convolutional layer contains 128 non-zero weights, corresponding to 128 1×1 convolutions. Therefore, the number of multiplication operations in the initial convolutional layer is 128×128=16384, and the number of addition operations is (128-32)×128=12288. After quantizing the initial convolutional layer to 1.58-bit precision, as shown in Table \ref{tab:1}, the number of non-zero weights is reduced to 8 and only 4 output channels remain non-zero. Accordingly, the number of Boolean operations is 8×128=1024, the number of additions is 4×128=512, and the number of indicator-function evaluations is 4×128=512.

\begin{table}[htbp]
  \centering
  \caption{Non-zero Weights in the Quantized Initial Conv0}
  \begin{tabular}{c *{4}{>{\centering\arraybackslash}m{1.2cm}}} 
    \toprule
    \diagbox[height=0.8cm]{Input Channel}{Output Channel} & 0 & 1 & 2 & 3 \\
    \midrule
    1  & -1 & 0  & 1  & 0  \\
    15 & 1  & 0  & -1 & 0  \\
    24 & 0  & 1  & 0  & -1 \\
    25 & 0  & -1 & 0  & 1  \\
    \bottomrule
  \end{tabular}
  \label{tab:1}
\end{table}

Gohr ND uses a residual block as its feature-extraction module, which consists of two 3×3 convolutions (with 32 input channels and 32 output channels), batch-normalization layers, and ReLU activation functions. According to Equation (\ref{eq:6}) and Equation (\ref{eq:7}), the convolution computation can be transformed into Boolean operations. Table \ref{tab:2} presents the simplified Boolean expressions for several channels obtained from the first convolutional layer in the residual block, where the subscripts denote the input channels.As shown in Table \ref{tab:2}, taking output channel 0 as an example, the number of non-zero weights in the original Gohr ND is 32×9=288. Accordingly, the number of multiplication operations is 288×128=36864, and the number of addition operations is (288-1)×128=36736. In contrast, after lightweighting, output channel 0 contains only 21 non-zero weights, resulting in 21×128=2688 Boolean operations, (21-1)×128=2560 additions, and 32 indicator-function evaluations.

\begin{table}[htbp]
  \centering
  \caption{Boolean Expressions Corresponding to Conv1 (Partial)}
  \begin{tabular}{c >{\centering\arraybackslash}m{14cm}} 
    \toprule
    Channel & Equation \\
    \midrule
    \multirow{3}{*}{0} 
    & $\displaystyle \text{Sum1} = H_1 + I_1 + H_{15} + I_{15} + B_{24} + D_{24} + E_{24} + B_{25} + C_{25} + D_{25} + E_{25}$ \\
    & $\displaystyle \text{Sum2} = B_1 + D_1 + E_1 + F_1 + G_1 + B_{15} + C_{15} + E_{15} + F_{15} + G_{15}$ \\
    & $\displaystyle Y = \mathbb{I}(\text{Sum1} > \text{Sum2})$ \\
    \midrule
    \multirow{3}{*}{1}
    & $\displaystyle \text{Sum1} = I_1 + F_1 + I_{15} + F_{15} + A_{24} + D_{24} + E_{24} + A_{25} + D_{25} + E_{25}$ \\
    & $\displaystyle \text{Sum2} = A_1 + D_1 + E_1 + H_1 + B_7 + A_{15} + D_{15} + E_{15} + H_{15} + F_{24} + G_{24} + F_{25} + I_{25}$ \\
    & $\displaystyle Y = \mathbb{I}(\text{Sum1} > \text{Sum2})$ \\
    \midrule
    \multirow{3}{*}{2}
    & $\displaystyle \text{Sum1} = C_1 + F_1 + H_1 + C_{15} + F_{15} + H_{15} + B_{24} + I_{24} + B_{25} + I_{25}$ \\
    & $\displaystyle \text{Sum2} = B_1 + E_1 + I_1 + B_7 + B_{15} + D_{15} + E_{15} + I_{15} + C_{24} + E_{24} + F_{24} + C_{25} + E_{25} + F_{25}$ \\
    & $\displaystyle Y = \mathbb{I}(\text{Sum1} > \text{Sum2})$ \\
    \midrule
    \multirow{3}{*}{3}
    & $\displaystyle \text{Sum1} = D_1 + E_1 + I_1 + D_{15} + E_{15} + I_{15} + B_{24} + H_{24} + B_{25}$ \\
    & $\displaystyle \text{Sum2} = B_1 + F_1 + G_1 + H_1 + B_{15} + G_{15} + H_{15} + E_{24} + E_{25}$ \\
    & $\displaystyle Y = \mathbb{I}(\text{Sum1} > \text{Sum2})$ \\
    \bottomrule
  \end{tabular}
  \label{tab:2}
\end{table}

In the feature-extraction module of the original Gohr ND, a single convolution kernel has 32 input channels and 32 output channels, with a total of 32×32×9=9216 weights. The number of multiplications is computed as the product of the number of weights and the output feature dimension, i.e., 9216×128=1179648, while the number of additions is (9216-32)×128=1175552. Since the feature-extraction module consists of two 3×3 convolutional layers, the total operation counts are twice those of a single convolutional layer, yielding 1179648×2=2359296 multiplications and 1175552×2=2351104 additions. For the lightweighted ND, the numbers of non-zero weights are 671 and 2081 for the first and second convolutional kernels, respectively. Accordingly, the feature-extraction module requires 352256 Boolean operations, 344064 additions, and 8192 indicator-function evaluations.

For the prediction head of the ND, its structure consists of one fully connected layer with an input of 4096 channels and an output of 64 channels, and another fully connected layer with an input of 64 channels and an output of 64 channels. These two fully connected layers can be transformed into expressions of the same form as Equation (\ref{eq:7}). The original Gohr ND prediction head contains 266240 multiplication operations and 266112 addition operations, while the lightweight ND prediction head contains 13877 Boolean operations, 13778 addition operations, and 128 indicator functions.

For the output layer of the ND, its structure is a fully connected layer with an input of 64 channels and an output of 2 channels. By examining its weights, it can be seen that the weights at corresponding positions of the two channels are opposite values, as shown in Table \ref{tab:3}. Accordingly, the two channels can be simplified to a single channel, and the output of this layer can be computed using Equation (\ref{eq:7}). The original Gohr ND output layer contains 128 multiplication operations and 126 addition operations, while the lightweight ND output layer contains 64 Boolean operations, 63 addition operations, and 1 indicator function.

\begin{table}[htbp]
  \centering
  \caption{Output Layer Weights (Partial)}
  \begin{tabular}{c *{8}{>{\centering\arraybackslash}m{0.8cm}}}
    \toprule
    \diagbox[height=0.8cm]{Input Channel}{Output Channel} & 0 & 1 & 2 & 3 & 4 & 5 & 6 & 7 \\
    \midrule
    0 & -1 & -1 & 1  &  1 &  1 & -1 & -1 &  1 \\
    1 &  1 & 1  & -1 & -1 & -1 &  1 &  1 & -1 \\
    \bottomrule
  \end{tabular}
  \label{tab:3}
\end{table}

A comparison of the classification accuracy before and after lightweighting is shown in Table \ref{tab:4}. The classification accuracy of the original Gohr ND is 94.95\%, while that of the lightweight ND is 92.21\%, and the simplification method proposed in this paper only reduces the model accuracy by 2.87\%. Since the lightweight ND only involves the summation of non-zero weights and uses Boolean operations instead of multiplication operations, the original model is counted in terms of the number of multiplication operations and addition operations, and the lightweight ND is counted in terms of the number of Boolean operations, addition operations and indicator functions, with the results shown in Table 4. The total number of operations of the lightweight ND is the sum of the number of Boolean operations, the number of addition operations and the number of indicator functions, and the total number of operations of the original ND is the sum of the number of multiplication operations and the number of addition operations. Calculations show that the total number of operations of the lightweight ND is 13.9\% of that of the original ND.

\begin{table}[htbp]
  \centering
  \caption{Comparison of the ND Before and After Lightweighting}
  \begin{tabular}{c *{5}{>{\centering\arraybackslash}m{2.5cm}}}
    \toprule
    Model & Number of Multiplication/Boolean Operations & Number of Addition Operations & Number of Indicator Functions & Accuracy \\
    \midrule
    Gohr Distinguisher & 2642048 & 2629630 & 0  &  94.95\%  \\
    Lightweight Distinguisher &  367221 & 358417  & 8833 & 92.21\%  \\
    \bottomrule
  \end{tabular}
  \label{tab:4}
\end{table}

After lightweighting the initial convolutional layer of Gohr model using the method proposed in this paper, the resulting 1.58‑bit convolutional layer exhibits extremely high sparsity. The original initial convolutional layer consists of 1×1 convolution, normalization, and the ReLU nonlinear activation function, with 32 output channels and 4 input channels, whereas the lightweighted 1.58‑bit convolutional layer has non-zero weights only in the four output channels 1, 15, 24, and 25, with all other output channels being fully zero channels. We can construct the corresponding input–output truth table for the input $(C_l, C_r, C_l', C_r')$ of the initial convolutional layer over the non-zero channels, thereby deriving the Boolean operation expressions corresponding to the 1.58‑bit initial convolutional layer to replace this convolutional layer, whose Boolean expressions are given in Table \ref{tab:5}. Furthermore, this paper compares the classification accuracy before and after simplifying this layer, and the results are shown in Table \ref{tab:6}, from which it can be seen that the simplification of this convolutional layer only leads to a classification accuracy loss of 0.3\%, while the computational complexity is significantly reduced.

\begin{table}[htbp]
  \centering
  \caption{Boolean Expressions Corresponding to the Conv0}
  \begin{tabular}{>{\centering\arraybackslash}m{3cm} >{\centering\arraybackslash}m{6cm}} 
    \toprule
    Channel & Boolean Expressions \\
    \midrule
    1  & $\displaystyle C_l' \land \overline{C_l}$ \\   
    15 & $\displaystyle C_l \land \overline{C_l'}$ \\  
    24 & $\displaystyle C_r \land \overline{C_r'}$ \\  
    25 & $\displaystyle C_r' \land \overline{C_r}$ \\   
    \bottomrule
  \end{tabular}
  \label{tab:5} 
\end{table}

\begin{table}[htbp]
  \centering
  \caption{Comparison of Training Accuracy of the ND Before and After Lightweighting the Conv0}
  \begin{tabular}{>{\centering\arraybackslash}m{4cm} >{\centering\arraybackslash}m{4cm}>{\centering\arraybackslash}m{4cm}}
    \toprule
    After Lightweighting & Before Lightweighting & accuracy loss \\
    \midrule
    94.64\%  & 94.95\% & 0.3\% \\
    \bottomrule
  \end{tabular}
  \label{tab:6}
\end{table}

\section{\quad Conclusion}
Aiming at the problem that ND usually involve a large number of 32-bit multiplications with high complexity and potential redundancy, this paper proposes a lightweight method for ND based on quantization-aware training. This method adopts the quantization-aware training scheme and quantizes the weights of the ND to 1.58-bit \{0, ±1\} via learnable step size quantization, replaces the multiplication operations in convolution computations with Boolean operations, and substitutes the ReLU nonlinear activation functions in the original model with indicator functions for magnitude comparison, so that the original ND involving 32-bit multiplications is finally transformed into a lightweight structure only consisting of Boolean operations, addition operations and indicator functions. Experimental results show that the lightweight ND contains more zero-valued weights, its total number of operations is 13.9\% of that of the original ND, which reduces the computational overhead, and the lightweight ND only decreases the classification accuracy by 2.87\%. When the proposed method only applies lightweight processing to the initial convolutional layer, the 1×1 convolutional layer with 4 input channels and 32 output channels can be replaced by 4 Boolean operations with a 16-bit sequence length, while the classification accuracy is reduced by only 0.3\%.

\section{\quad Acknowledgments}
The authors would like to thank Professor Cui Ting from the Information Engineering University for his guidance on the writing and revision of this paper.
This work was supported by the National Natural Science Foundation of China (Grant No. 62271504).

\section*{Funding}
National Natural Science Foundation of China (Grant No. 62271504)

\renewcommand{\refname}{References}

\end{document}